# ATTENTIVE BATCH NORMALIZATION FOR LSTM-BASED ACOUSTIC MODELING OF SPEECH RECOGNITION

*Fenglin Ding, Wu Guo, Lirong Dai, Jun Du*

National Engineering Laboratory for Speech and Language Information Processing
University of Science and Technology of China, Hefei, China
flding@mail.ustc.edu.cn, {guowu,lrdai,jundu}@ustc.edu.cn

## ABSTRACT

Batch normalization (BN) is an effective method to accelerate model training and improve the generalization performance of neural networks. In this paper, we propose an improved batch normalization technique called attentive batch normalization (ABN) in Long Short Term Memory (LSTM) based acoustic modeling for automatic speech recognition (ASR). In the proposed method, an auxiliary network is used to dynamically generate the scaling and shifting parameters in batch normalization, and attention mechanisms are introduced to improve their regularized performance. Furthermore, two schemes, frame-level and utterance-level ABN, are investigated. We evaluate our proposed methods on Mandarin and Uyghur ASR tasks, respectively. The experimental results show that the proposed ABN greatly improves the performance of batch normalization in terms of transcription accuracy for both languages.

## 1. INTRODUCTION

The application of deep neural networks (DNNs) to automatic speech recognition (ASR) has achieved tremendous success due to its superior performance over traditional GMM/HMM acoustic models [1, 2]. Recurrent neural networks (RNNs), which can model the temporal dependencies of speech signals, are the state of the art techniques for ASR. However, standard RNN training suffers from the gradient vanishing and exploding problem [3]. To address this problem, long short-term memory (LSTM) has been proposed [4] and achieved great success in ASR [5-8]. However, LSTM is computationally expensive to train and difficult to parallelize. As a result, many techniques have been proposed to enhance the performance of LSTM with respect to both its speed and generalization. In particular, batch normalization (BN) [9], which uses mini-batch statistics to standardize features, has significantly reduced the training time and improved the generalization of neural networks.

As mentioned in [9], the BN transformation may not be appropriate or work well in conjunction with the nonlinear transformation that follows. Therefore, the linear transformation with scaling and shifting parameters is introduced to adjust the BN transformation. It is critical to guarantee the performance of the linear transformation in batch normalization. This paper focuses on improving the performance of batch normalization by learning more helpful normalization parameters in LSTM-based acoustic modeling.

In addition, researchers have investigated improving layer normalization by dynamically generating the scaling and shifting parameters [10]. Motivated by this, we explore how to dynamically generate the normalization parameters in BN. We propose two novel strategies to generate parameters: at the frame-level and at the utterance-level. Furthermore, attention mechanisms are introduced into our proposed BN to focus on the context-discriminative deep features, and we call it attentive batch normalization (ABN). The scaling and shifting parameters generation network is jointly trained and optimized with the main network. Experiments are conducted on the speech recognition tasks of two very different languages: Mandarin and Uyghur, respectively. The results show that our proposed methods greatly improve the performance of batch normalization for both languages.

The rest of this paper is organized as follows. First, the standard batch normalization method will be reviewed in section 2. Section 3 provides a detailed description of our proposed parameters generation techniques. Section 4 shows the experimental setup and provides other details, including the performance and discussions, and section 5 concludes the paper.

## 2. RELATED WORK

### 2.1. Batch Normalization

Batch normalization [10] is originally designed to alleviate the issue of internal covariate shifting, which is a common problem in DNN training. It can accelerate the DNN training and has been shown to be effective in various applications. From a mathematical perspective, BN normalizes the activations of a deep network within a mini-batch. For a neural activation $x$ of dimension $p$, $x = \{x^{(1)},...,x^{(i)},...,x^{(p)}\}$, each dimension is normalized by

$$\hat{x}^{(i)} = \frac{x^{(i)} - \mu_B}{\sqrt{\sigma_B^2 + \varepsilon}} \quad (2)$$

where $\varepsilon$ is a small positive constant to prevent numerical instability, $\mu_B$ and $\sigma_B^2$ are the mini-batch mean and variance.

However, normalizing the intermediate activations reduces the representational power of the layer. To address this issue, BN introduces additional learnable parameters $\gamma$ and $\beta$, which scale and shift the data $\hat{x}^{(i)}$, respectively.

$$y^{(i)} = \gamma^{(i)} \hat{x}^{(i)} + \beta^{(i)}$$

where $\gamma$ and $\beta$ are parameters to be trained along with the original model parameters.

## 2.2. Batch Normalization in LSTM

For a standard feedforward layer in a neural network,
$$\mathbf{y} = \phi(\mathbf{Wx} + \mathbf{b}) \quad (5)$$
where $\mathbf{W}$ is the weights matrix, $\mathbf{b}$ is the bias vector, $\mathbf{x}$ is the input of the layer and $\varphi$ is an activation function. BN can be applied as follows:
$$\mathbf{y} = \phi(BN(\mathbf{Wx})) \quad (6)$$
Note that the bias vector has been removed since its effect is canceled by the standardization.

However, it is more difficult to apply BN in a recurrent architecture. Researchers have investigated many algorithms to apply BN on the recurrent neural network [11, 12, 13, 14]. In this paper, we also applied BN to the input-to-hidden transitions of the LSTMs. However, different from the operation in [11], BN is applied to LSTM before the weights matrix as follows:

$$\mathbf{i}_t = \text{sigmoid}(\mathbf{W}_{hi}\mathbf{h}_{t-1} + \mathbf{W}_{xi}BN(\mathbf{x}_t))$$
$$\mathbf{f}_t = \text{sigmoid}(\mathbf{W}_{hf}\mathbf{h}_{t-1} + \mathbf{W}_{xf}BN(\mathbf{x}_t))$$
$$\mathbf{c}_t = \mathbf{f}_t \odot \mathbf{c}_{t-1} + \mathbf{i}_t \odot \tanh(\mathbf{W}_{hc}\mathbf{h}_{t-1} + \mathbf{W}_{xc}BN(\mathbf{x}_t)) \quad (7)$$
$$\mathbf{o}_t = \text{sigmoid}(\mathbf{W}_{ho}\mathbf{h}_{t-1} + \mathbf{W}_{xo}BN(\mathbf{x}_t) + \mathbf{W}_{co}\mathbf{c}_t)$$
$$\mathbf{h}_t = \mathbf{o}_t \odot \tanh(\mathbf{c}_t)$$

where $\mathbf{W}_{h\cdot}$ are the recurrent weight matrices and $\mathbf{W}_{x\cdot}$ are the input-to-hidden weight matrices, $\mathbf{i}_t$, $\mathbf{f}_t$ and $\mathbf{o}_t$ are respectively the input, forget and output gates, and $\mathbf{c}_t$ is the cell. Sigmoid( ) is the logistic sigmoid function, and tanh is the hyperbolic tangent function.

For a bidirectional LSTM framework, BN is simultaneously applied to the forward and backward LSTMs.

## 3. ATTENTIVE BATCH NORMALIZATION

The proposed ABN can be categorized into the frame-level and utterance-level, respectively. For the frame-level scheme, the scaling and shifting parameters are generated for a whole utterance as with the standard BN. Additionally, for the utterance-level scheme, the normalization parameters are generated for each frame of the input utterance. Furthermore, different from previous work, we use the normalized hidden activations to generate the scaling and shifting parameters.

Figure 1 shows the architecture of deep LSTM with ABN. An auxiliary extractor is added to the main LSTM to dynamically generate the parameters of BN for that layer. More details are discussed below.

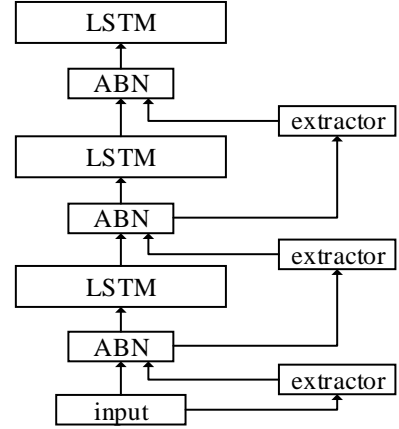

**Fig. 1.** The architecture of our proposed ABN applied to deep LSTM.

### 3.1. Frame-level ABN

Suppose that $\mathbf{h}_t^{l-1}$ denotes the hidden activation of the $l$-1$^{th}$ layer at time step $t$, which is also the input of the $l^{th}$ layer. $\tilde{\mathbf{h}}_t^{l-1} = BN(\mathbf{h}_t^{l-1})$ denotes the normalized hidden activation. For the normalization parameters generation network, a nonlinear transformation is first applied to project the hidden activation to a low dimension space to accelerate the computation as follows:

$$\mathbf{e}_t^{l-1} = \tanh(\mathbf{W}_e \tilde{\mathbf{h}}_t^{l-1} + \mathbf{b}_e) \quad (8)$$

where $\mathbf{W}_e$ is a $d_e \times p$ weight matrix and $d_e$ is set to be less than $p$.

We use the weighted summation of each frame to generate the normalized parameters. The mean of all the elements in $\mathbf{e}_t^{l-1}$ can measure the importance of the $t^{th}$ frame-level representation. With the softmax function, the attention weight for each frame of the utterance can be calculated as

$$\alpha_t = \frac{\exp(\text{mean}(\mathbf{e}_t^{l-1}))}{\sum_\tau \exp(\text{mean}(\mathbf{e}_\tau^{l-1}))} \quad (9)$$

The weighted mean of the whole utterance can be easily computed with $\alpha_t$ serving as the combination weights.

$$\mathbf{u} = \sum_t \alpha_t \mathbf{e}_t^{l-1} \quad (10)$$

Finally, the scaling and shifting parameters are generated as a linear transformation of the weighted mean:

$$\boldsymbol{\gamma}^l = \mathbf{W}_\gamma^l \mathbf{u} + \mathbf{b}_\gamma^l$$
$$\boldsymbol{\beta}^l = \mathbf{W}_\beta^l \mathbf{u} + \mathbf{b}_\beta^l \quad (11)$$

### 3.2. Utterance-level ABN

In this scheme, we use a self-attention mechanism [15] to adjust the weights of different frames. We follow the standard formation of the scaled dot-product attention that was proposed in [15]. For the normalization parameters generation network, the key, value and query are the projections of $\tilde{\mathbf{h}}_t^{l-1}$ in the $d_a$ dimensional space:

$$\mathbf{K}_t = \mathbf{W}_k * \tilde{\mathbf{h}}_t^{l-1}$$
$$\mathbf{Q}_t = \mathbf{W}_q * \tilde{\mathbf{h}}_t^{l-1} \quad (12)$$
$$\mathbf{V}_t = \mathbf{W}_v * \tilde{\mathbf{h}}_t^{l-1}$$

where $\mathbf{W}_k$, $\mathbf{W}_q$ and $\mathbf{W}_v$ are $d_a \times p$ projection matrices, and $d_a$ denotes the dimension of the different keys, queries and values.

The attention weight $\alpha_t$ of each current frame $\tilde{\mathbf{h}}_t$ against all the context deep features in the whole utterance, i.e., $\{\tilde{\mathbf{h}}_1,...,\tilde{\mathbf{h}}_t,...,\tilde{\mathbf{h}}_T\}$, is computed by normalizing the similarity scores $e_{t,\tau}$ between the query $\mathbf{Q}_t$ and each key $\mathbf{K}_\tau$ as follows:

$$\alpha_{t,\tau} = \frac{\exp(e_{t,\tau})}{\sum_{\tau'=1}^{T} \exp(e_{t,\tau'})} \quad (13)$$

$$e_{t,\tau} = \frac{\mathbf{K}_\tau^T \mathbf{Q}_t}{\sqrt{d_a}} \quad (14)$$

A context vector $\mathbf{c}_t$ is formed at each time $t$ as a weighted sum of the projected features $\mathbf{V}_t$ with the attention vector $\alpha_t$ serving as the combination weights:

$$\mathbf{c}_t = \sum_{\tau=1}^{T} \alpha_{t,\tau} \mathbf{V}_t \quad (15)$$

Then, $\mathbf{c}_t$ is used to finally generate the scaling and shifting parameters for the features at time step t:

$$\boldsymbol{\gamma}_t^l = \mathbf{W}_\gamma^l \mathbf{c}_t + \mathbf{b}_\gamma^l$$
$$\boldsymbol{\beta}_t^l = \mathbf{W}_\beta^l \mathbf{c}_t + \mathbf{b}_\beta^l \quad (16)$$

This utterance-level design expects to provide a more discriminative scaling and shifting transformation for the hidden activation. The network can focus on the context-discriminative deep features, which is beneficial to speech recognition.

Briefly, by dynamically generating the scaling and shifting parameters in BN, the training can avoid getting a poor choice of the scale and center due to random initialization. Furthermore, the introduction of attention mechanisms make the whole transformation focus on the context-discriminative deep features in the acoustic modeling. This approach can both alleviate the issue of the internal covariate shifting in the deep network and normalize the model for acoustic variability.

## 4. EXPERIMENTS

### 4.1. Database

The experiments are conducted on the recognition tasks of two languages, Mandarin and Uyghur. The open-source speech corpus AISHELL-1 [16] is used for the Mandarin speech recognition. All the speech files are sampled at 16 K Hz with 16 bits. The training set contains 150 hours of speech (120,098 utterances) recorded by 340 speakers. The development set contains 20 hours of speech (14,326 utterances) recorded by 40 speakers. In addition, the test set contains 10 hours of speech (7,176 utterances) recorded by 20 speakers. The speakers of the training set, development set, and test set are not overlapped.

For the Uyghur speech recognition, we use a Uyghur conversational telephone speech corpora collected by the Speechocean Corporation, King-ASR-450. All the speech files are sampled at 8 K Hz with 16 bits. The King-ASR-450 corpus contains 120 hours of (79,149 utterances) spontaneous dialog speech data that were recorded by 200 speakers. We randomly select 2,000 utterances (3.15-hour) from 10 speakers as the test set and 1,000 utterances (1.58-hour) from the remaining 190 speakers as the development set. The remaining utterances of the 190 speakers are used for training.

### 4.2. Model Setup

We use connectionist temporal classification (CTC) [17] based speech recognition systems in our experiments. For both systems, the input acoustic features are 108-dimensional filter-bank features (36 filter-bank features, delta coefficients, and delta-delta coefficients) with mean and variance normalization. All neural acoustic models in the experiments have three bidirectional LSTM hidden layers with 512 LSTM cells. To improve the recognition performance and training efficiency, we append two convolutional neural network (CNN) layers before the LSTM layers. For the baseline, the bottom two layers are 2D convolution layers with output channels of 64 and 256. Each convolution layer is followed by a max-pooling layer with a stride of 2 in the time dimension for finally down sampling utterances to a quarter of their original lengths. Standard batch normalization is applied to all LSTM layers, as in Eq. (1). We use a dropout rate of 0.3 for the LSTM layers to avoid overfitting.

According to the statistical information of the transcripts, we collect 4294 Chinese characters in the training and development sets for the Mandarin case. With the special symbol (blank) involved, 4295 modeling graphemes are used as the output units. Meanwhile, for the output of the Uyghur acoustic model, we use the byte pair encoding (BPE) [18] algorithm to generate the modeling units. We follow the operation of generating sub-words using BPE algorithms that was proposed in [19]. Uyghur is written in the Latin alphabet. Taking into account the position information, 1859 units, i.e., 1858 BPE sub-words and a blank, are used for the output layer.

In the proposed ABN, the size of the parameters generation network, i.e., $d_e$ and $d_a$, is set to 256 to speed up the training. The normalized hidden activation of the previous layer is used to generate the scaling and shifting parameters for the current layer. We also use dropout for the parameters generation network to improve performance.

### 4.3. Training and Decoding

We sort all the utterances in the training set in descending order and set a maximum number of frames of 5000 to control the batch size. The number of utterances included in each mini-batch is $5000/L_{max}$, where / means the rounding operation and $L_{max}$ denotes the length of the longest utterance of the mini-batch. All utterances whose lengths are less than $L_{max}$ are unified by using a zero-padding operation. Therefore, the size of each mini-batch is variable but will not exceed 5000.

The network is trained to minimize the CTC loss function with an initial learning rate of 0.0001. The development set is used for learning rate scheduling and early stopping. We start to halve the learning-rate when the relative improvement falls below 0.004, and the training ends if the relative improvement is lower than 0.0005.

PyTorch toolkits [20] are used in our model training process. All the model parameters are randomly initialized and updated by Adam [21]. The trigram language model, which is trained by using the transcription of the training set, is used in the decoding procedure.

### 4.4. Results and Discussion

| Model | Model Size(M) | Dev. | Test |
|---|---|---|---|
| BN | 85.85 | 8.35 | 9.71 |
| ABN-F | 91.98 | 7.52 | 8.43 |
| ABN-U | 95.06 | 7.40 | 8.40 |

**Table 1.** The CERs (%) of the development and test sets of AISHELL-1.

The character error rate (CER) is used as an evaluation criterion for Mandarin recognition. Table 1 shows the results of standard BN and our proposed ABN on AISHELL-1. ABN-F and ABN-U stand for the frame-level and utterance-level ABN, respectively. As shown in Table 1, the model trained with ABN outperforms the baseline model for both of the development and test sets. The two proposed ABN schemes both greatly improve the performance over standard batch normalization.

| Model | Model Size(M) | Dev. | Test |
|---|---|---|---|
| BN | 76.10 | 10.33 | 10.36 |
| ABN-F | 82.23 | 9.43 | 9.31 |
| ABN-U | 85.31 | 9.65 | 9.61 |

**Table 2.** The WERs (%) of development and test sets of King-ASR-450

The word error rate (WER) is used as an evaluation criterion for Uyghur recognition. The experimental results of standard BN and the proposed ABN on King-ASR-450 are shown in Table 2. Compared with Mandarin, the Uyghur datasets have much fewer speech utterances and speakers. However, the proposed strategies also outperform standard batch normalization for both of the development and test sets. However, ABN-U produces worse performance than the ABN-F.

From the above two group of experiments, the utterance-level ABN does not provide better performance than the frame-level ABN. It is suspected that the more complex structure of the parameters generation network aggravates the overfitting of the network, especially in smaller training sets, such as those in the Uyghur experiments.

### 5. CONCLUSIONS

In this paper, we propose an improved batch normalization technique called attentive batch normalization (ABN) in LSTM-based acoustic modeling for speech recognition. We propose to generate the normalization parameters at the frame-level and at the utterance-level. Experiments are carried out by using a large vocabulary language, Mandarin, and a low resource language, Uyghur. The results show that our proposed methods greatly improve the performance of batch normalization for both languages. In future work, we plan to simplify the parameters extraction in the utterance-level ABN to further improve the performance.

### 6. ACKNOWLEDGEMENTS

This work was partially funded by the National Key Research and Development Program of China (Grant No. 2016YFB100 1303) and the National Natural Science Foundation of China (Grant No. U1836219).

### 7. REFERENCES

[1] G. Hinton, L. Deng, D. Yu, et al., "Deep neural networks for acoustic modeling in speech recognition: